\documentclass{aa}

\usepackage{txfonts}

\usepackage{graphicx}

\begin{document}

\def\P{\bar{\Phi}}

\def\st{\sigma_{\rm T}}

\def\vk{v_{\rm K}}

\def\sles{\lower2pt\hbox{$\buildrel {\scriptstyle <}
   \over {\scriptstyle\sim}$}}

\def\sgreat{\lower2pt\hbox{$\buildrel {\scriptstyle >}
   \over {\scriptstyle\sim}$}}

\title{The Coughing Pulsar Magnetosphere}

\author{Ioannis Contopoulos
\thanks{\emph{Present address:}Research Center for Astronomy 
\& Applied Mathematics, Academy of Athens, 
\email{icontop@academyofathens.gr}}}

\institute{200 Akti Themistokleous Str., Piraeus 18539, Greece}


\date{Received 28 March 2005 / Accepted 28 June 2005}

\abstract{
Polar magnetospheric gaps consume a fraction of the electric
potential that develops 
accross open field lines. This effect modifies significantly
the structure of the axisymmetric pulsar magnetosphere. 
We present numerical
steady-state solutions for various values of the 
gap potential.
We show that a charge starved magnetosphere contains significantly
less electric current than one with freely 
available electric charges. 
As a result, electromagnetic neutron star braking becomes
inefficient. 
We argue that the magnetosphere may spontaneously rearrange
itself to a lower energy configuration through
a dramatic release of electromagnetic field energy
and magnetic flux. Our results might be relevant
in understanding the recent December 27, 2004 burst observed in 
SGR~1806-20.


}

\maketitle

\section{Introduction}

The magnetosphere of a rotating neutron star with
polar magnetic field $B_*$, mass $M_*\sim M_\odot$, radius
$r_*\sim 10{\rm km}$, magnetic dipole moment $\mu=
r_*^3 B_*/2$, and angular velocity $\Omega$ is expected 
to radiate electromagnetic energy at a rate
\begin{equation}
L_{\rm em}\sim \alpha\frac{B_*^2 \Omega^4 r_*^6}{c^3}\ .
\label{1}
\end{equation}
$\alpha$ is a factor of order unity (\cite{Beskin97})\footnote{
$\alpha=\frac{1}{6}\sin^2\theta$ for a misalligned dipole rotating in vacuum.
In that scenario, an alligned magnetic rotator
($\theta\approx 0$) does not radiate. However,
when the neutron star is not surrounded by vacuum, one needs
to consider the structure of its rotating charged relativistic 
Goldreich-Julian-type magnetosphere (Goldreich \& Julian~1969). 
In that case, the electric currents that flow through the magnetosphere
lead to electromagnetic energy losses comparable to the
ones for a misalligned magnetic rotator.
See the Appendix for a general calculation.}.
The source of the radiation is the neutron star rotational
kinetic energy which is lost at a rate
\begin{equation}
L_{\rm kinetic}\sim \frac{4}{5}M_*r_*^2 \Omega\dot{\Omega}\ .
\label{2}
\end{equation}
It is usual to equate eqs.~\ref{1} \& \ref{2} and thus obtain
an estimate of the stellar magnetic field $B_*$.
In general, however, the two do not have to be equal.
As we shall see below, in the case of axisymmetry,
electromagnetic torques need to be significantly revised.

In the context of ideal axisymmetric MHD, 
electric charges are available
in copious amounts and move freely along magnetic field lines,
shorting out any component of the electric field that might
arise along
the magnetic field. As a result, magnetic flux surfaces become
equipotentials, and an electric field ${\bf E}$ develops
accross magnetic field lines (${\bf E}\cdot{\bf B}=0$)
with magnitude
\begin{equation}
E=\frac{r\Omega_F}{c}B_p\ ,
\end{equation}
where, $\Omega_F$, a constant along magnetic flux surfaces
(see below), can be thought of as the
angular velocity of rotation of magnetic field lines ($r$ is
the cylindrical radius; $B_p$ is
the poloidal component of the magnetic field).
The source of the electric potential accross magnetic
field lines is obviously the rotating magnetized conducting
surface of the neutron star which acts as a unipolar inductor.
For the sake of simplicity, most studies of the axisymmetric
pulsar magnetosphere have assumed that the full potential drop
induced accross field lines along the surface
of the star continues to manifest itself all along those
field lines, i.e. $\Omega_F=\Omega$. It has been pointed out, however, 
(e.g. Ruderman \& Sutherland, 1975) that 
`open magnetic field lines play a role analogous to that
of conducting wires in ordinary circuits. If the wire is broken
near the pulsar surface, a potential drop develops accross the gap'.
The presence
of such gaps obviously reduces the electric potential available
accross open field lines, and thus the electromagnetic energy
power radiated at large distances.
Models of particle acceleration and pair creation
above the polar cap of rotation-powered pulsars yield 
potential drops near the surface of the star
of the order of $10^{12}$~Volts (e.g. \cite{HArons01}; Arons
personal communication),
and therefore, in general, $\Omega_F < \Omega$.

Beskin \& Malyshkin 1998  
took the above well known effect into account
in their calculation of the modified magnetospheric structure
inside the light cylinder.
In the present paper we obtain the first global 
solution of this problem.
In \S~2 we formulate the problem and the numerical method
that we implement for its solution.
In \S~3 we obtain the structure of the magnetosphere
for various values of $\Omega_F$ in the range $[0,\Omega]$
and argue that the magnetosphere 
may switch between solutions, releasing
energy in the process.
In \S~4 we discuss the relevance of our results in
understanding the recent December 27, 2004 SGR-1806-20 burst.
Our conclusions are summarized in \S~5.

\section{The differentially rotating magnetosphere}

We will work in cylindrical spatial coordinates $r,\phi,z$,
and will consider only the axisymmetric case where
the magnetic dipole axis is aligned with the axis of rotation.
This simplification allows us
to introduce the magnetic flux function $\psi$
($\psi/r$ is the $\phi$-component of the magnetic
vector potential), 
the poloidal electric current function $A=A(\psi)$
(the poloidal electric current contained within
the magnetic flux surface $\psi$ is equal to $Ac/2$; $B_\phi=A/r$), 
and the magnetic field line
`rotational velocity' $\Omega_F=\Omega_F(\psi)$. The various 
magnetospheric physical quantities are obtained as follows:
\begin{equation}
{\bf B}=\frac{1}{r}\left(-\psi_z,A,\psi_r\right)\ ,
\end{equation}
\begin{equation}
{\bf E}=-\frac{\Omega_F}{c}\nabla\psi=
-\frac{\Omega_F}{c}\left(\psi_r,0,\psi_z\right)\ ,
\label{Efield}
\end{equation}
\begin{equation}
{\bf J}=\frac{c}{4\pi}\nabla\times {\bf B}
=-\frac{c}{4\pi r}\left(A'\psi_z,
\psi_{rr}-\frac{\psi_r}{r}+\psi_{zz},
-A'\psi_r\right)\ ,
\end{equation}
\begin{equation}
\rho_e=\frac{1}{4\pi}\nabla\cdot{\bf E}=
-\frac{\Omega_F}{4\pi c}\left(\psi_{rr}+\frac{\psi_r}{r}
+\psi_{zz}\right)
-\frac{{\Omega}'_F}{4\pi c}(\nabla\psi)^2\ .
\end{equation}
Here, and in what follows, 
$\psi_x\equiv \partial \psi /\partial x$.
Also, $(\ldots)'\equiv {\rm d}(\ldots)/{\rm d}\psi$.
When we neglect inertia, force balance requires that
\begin{equation}
\frac{1}{c}{\bf J}\times {\bf B}+\rho_e{\bf E}=0\ .
\label{ff1}
\end{equation}
Following Gruzinov~2005, we take
\begin{equation}
c=\mu=\Omega=1\ ,
\end{equation}
and thus eq.~\ref{ff1} becomes
\begin{equation}
(1-r^2\Omega_F^2)\left(\psi_{rr}+\frac{\psi_r}{r}+\psi_{zz}\right)
-\frac{2\psi_r}{r}=-AA'
+r^2\Omega_F\Omega_F'(\nabla \psi)^2
\label{pulsareq}
\end{equation}
This is a more general form of the pulsar equation than
the one considered in Contopoulos, Kazanas \& Fendt (hereafter CKF)
where $\Omega_F \equiv\Omega=1$ everywhere. 

$\Omega_F$ is related to the magnetospheric
potential drop $V_F$ between
the axis and any magnetic flux surface $\psi$
(eq.~\ref{Efield}), namely
\begin{equation}
V_F(\psi) =\frac{1}{c}\int_0^\psi \Omega_F {\rm d}\psi
\end{equation}
(in units $B_* r_*^3 \Omega^2/2c^2$). This is in general
{\em different} from the stellar potential drop between the pole and 
the footpoint on the surface of the star
of the magnetic flux surface $\psi$, namely
\begin{equation}
V_*(\psi) = \psi/c\ .
\label{Vstar}
\end{equation}
The difference 
\begin{equation}
V(\psi)\equiv V_*-V_F=\frac{1}{c}\int_0^\psi
 (1-\Omega_F) {\rm d}\psi\ .
\label{V1}
\end{equation}
is just the particle acceleration gap potential which devolops
{\em along the magnetic field} near 
the footpoint of the magnetic flux surface (e.g. \cite{Beskin97}).
In the region of closed field lines (hereafter the `dead zone'), 
there is no particle flow, and
therefore there is no need for the formation of 
particle acceleration gaps. We can thus express
\begin{equation}
\Omega_F(\psi)=\left\{
\begin{array}{ll}
\Omega_{Fo}\leq 1 & \mbox{along open field lines}\ 
\psi\leq \psi_{\rm open} \\
1 & \mbox{in the `dead zone'}
\end{array}
\right.
\label{OmegaFo}
\end{equation}
$\Omega_{Fo}(\psi)$ is determined by the particle acceleration gap 
microphysics and by the supply of charge carriers
at the base of the magnetosphere (see discussion in \S~4). 
In our present discussion, 
$\Omega_{Fo}(\psi)$ is essentially a free function.
For the sake of simplicity, we take $\Omega_{Fo}=$~const., 
as in \cite{BeskinM}.
The magnetospheric gap potential is, therefore, given by
\begin{equation}
V(\psi)=\left\{
\begin{array}{ll}
(1-\Omega_{Fo})\psi
 & \mbox{along open field lines}\ 
\psi\leq \psi_{\rm open} \\
0 & \mbox{in the `dead zone'}
\end{array}
\right.
\label{V}
\end{equation}
$V$ is minimal at the center of the polar cap and increases $\propto r^2$
as we move away from the axis.

Observational manifestation of the differential magnetospheric
rotation is thought to be found in the sub-pulse slow drifts accross
the pulse profile in almost aligned pulsars (e.g. \cite {Rankin03}). 
Interpretation of such drifts remains still rather sketchy.
We speculate that the sub-pulses are
associated with the above mentioned magnetospheric gaps
present around the separatrix between open and closed
field lines where the need for electric charge carriers
is the greatest (as we discuss below, this is where
closes the electric circuit of the
poloidal electric current that flows through the polar cap).
In most cases with observed sub-pulse drifts
($\sim 1$~sec period pulsars) $\Omega_{Fo}$ is expected to
be much smaller than $\Omega$. These gaps are 
probably carried around the axis of rotation
by the `friction' between the differentially rotating
open and closed line regions, and thus
their observed angular velocity is found to be
close to $\Omega$.

As we mentioned before, solutions of the pulsar equation
exist only for the simplest case $\Omega_F=1$ and
$\Omega_F'=0$. Even
in that case, though, a strong mathematival singularity,
the so called `light cylinder'
\begin{equation}
r_{\rm lc}=1\ ,
\end{equation}
makes the problem non-trivial. 
Note that this is just the force-free Alfven surface,
and only very recently
has a numerical method been presented which allowed us
to obtain a `smooth' solution that fills all space (CKF).
The main features of that numerical solution
(further refined in Gruzinov~2005) are:
\begin{enumerate}
\item The region of open field lines, the so called `polar cap',
is {\em slightly larger} than
the region of static dipolar field lines which cross the equator
beyond the distance $r=1$, namely
\begin{equation}
\psi_{\rm open}=1.23
\end{equation} 
(present calculation\footnote{In CKF, with a much lower numerical
resolution,  we obtained a value of ~1.36. Gruzinov~2005 obtained
a value of 1.27 with a numerical resolution comparable to our present one.}). 
We remind the reader that $\psi_{\rm open}=1$ for a static dipole,
and therefore we see that 
rotation `pulls' dipolar magnetic field lines out.
\item The distribution of poloidal electric current along
the open field lines is {\em very close} to the one in the 
\cite{Michel74} relativistic split monopole solution, namely
\[
A_{\rm CKF}\simeq
\]
\begin{equation}
\left\{
\begin{array}{ll} A_{\rm Michel}\equiv
-\psi\left(2-\frac{\psi}{\psi_{\rm open}}\right)
& \mbox{along open field lines} \\
0 & \mbox{in the `dead zone'}
\end{array}
\right.
\label{Michel}
\end{equation}
The electric current distribution deviates slightly from the above
near $\psi\sim \psi_{\rm open}$ where field lines follow
the singular shape of the separatrix between the open and closed
line regions (see fig.~\ref{figure:6} below).
\item The return current of the above current distribution,
$A(\psi_{\rm open})\sim \psi_{\rm open}$, flows
along the separatrix.
This implies the presence of  magnetic and electric field 
discontinuities accross the separatrix.
\item In general, the equatorial extent $r_c$ of the `dead zone'
may be taken as a free parameter (see section~5, Appendix).
It is very natural, however, to assume that the `dead zone'
extends all the way to the light cylinder\footnote{
Gruzinov~2005 shows that
this solution requires infinite magnetic fields at the
point $r=1,z=0$ (in the limit of infinitesimal grid size). 
\cite{Uzdensky03} and \cite{Lyubarskii90} argue against
infinite fields and thus conclude 
that the dead zone should end at some finite distance inside
the light cylinder.}. 
\item Open field lines become {\em monopole-like} around and beyond
the light cylinder.
\item $|{\bf B}|>|{\bf E}|$ everywhere\footnote{
This observation counteracts criticism
that the assumptions of ideal MHD may break down beyond 
the light cylinder (\cite{Ogura03}; \cite{Spitkovsky04}).
We believe that the source of the opposite result presented
in \cite{Ogura03} (their fig.~5) is due to their 
numerical boundary condition, eq.~3.3 \& fig.~1, namely that
field lines become horizontal at large radial distances.}.
\end{enumerate}

We are now ready to address the physically more interesting
case $\Omega_F'\neq 0$, in the simplest possible case
where $\Omega_F=
\Omega_{Fo}={\rm const.}< 1$ in the open field line
region, and $\Omega_F=1$ in the dead zone (eq.~\ref{OmegaFo}).
When $\psi\leq \psi_{\rm open}$, we can rewrite eq.~\ref{pulsareq} in
the new spatial coordinates $x\equiv \Omega_{Fo}r$ and
$y\equiv\Omega_{Fo}z$,
\begin{equation}
(1-x^2)\left(\psi_{xx}+\frac{\psi_x}{x}+\psi_{yy}\right)
-\frac{2\psi_x}{x}=-\frac{AA'}{\Omega_{Fo}^2}\ .
\label{pulsareq2}
\end{equation}
Eq.~\ref{pulsareq2} is the same as our original equation in CKF.
We thus expect that solutions of eq.~\ref{pulsareq} will be very similar
to the ones obtained in CKF. We would like to emphasize the
following interesting features:
\begin{enumerate}
\item As in CKF, it is natural to assume that
the corrotating `dead zone' extends all the way to the
light cylinder distance, i.e.  $r_c=r_{\rm lc}=1$.
The real mathematical singularity, however, is not at the light
cylinder, but at a certain distance outside, the
`open field light cylinder' 
\begin{equation}
r_{\rm oflc}=\Omega_{Fo}^{-1}>1\ .
\end{equation}
This is where we will apply the numerical iteration routine
developed in CKF. 
\item We also expect $\psi_{\rm open} \sim 1$ as in previous solutions. 
\item As in CKF, we expect to encounter similar
magnetic and electric field discontinuities accross the separatrix
between open and closed field lines.
\item The r.h.s. of eq.~\ref{pulsareq2}
is obtained through a numerical iteration
along the open field light cylinder that guarantees smooth
crossing of the singularity. Based on our experience, we expect this
function to be very close to $-AA'_{\rm CKF}$.
Therefore, to a good approximation,
\begin{equation}
A\simeq \Omega_{Fo}A_{\rm CKF}\ ,
\end{equation}
i.e. $A\propto \Omega_{Fo}$.
Obviously, as $\Omega_{Fo}\rightarrow 0$, $A\rightarrow 0$. As
we will see, this result has very interesting implications
for the electromagnetic torques
on the surface of the neutron star.
\end{enumerate}
Eq.~\ref{pulsareq} is elliptic with mixed
boundary conditions inside and outside the open 
field light cylinder $r=r_{\rm oflc}$:
\begin{enumerate}
\item $\psi=0$ along $r=0$, and $\psi=\psi_{\rm open}$
along the equator beyond $r=1$ (Dirichlet boundary 
conditions)\footnote{
As is shown in the Appendix
we are in general allowed to arbitrarily choose
the equatorial extent $r_c$
of the closed line region. In that case, $\psi_{\rm open}$
is obtained as a solution of eq.~\ref{pulsareq} inside the
open field light cylinder, i.e. {\em it is not}
an extra free parameter 
(see Goodwin {\em et al.}~2004 for a different point of view).}. 
\item $\psi_z=0$ (i.e. $B_r=0$) along the equator in the closed
line region $r< 1$ (Newman boundary condition). 
\item $\psi_r=AA'/(2\Omega_{Fo})$ along the open field light cylinder
$r=r_{\rm oflc}$  (Newman boundary condition). 
\item Finally, as in CKF, 
boundary conditions at infinity are irrelevant as long
as we rescale our spatial coordinates to new ones that map the full
$(r,z)=([0,\infty],[0,\infty])$ space to our finite grid size
$(r_{\rm new},z_{\rm new})=([0,2],[0,1])$. Note that this is not
the case for other numerical schemes where the integration is
constrained within finite spatial extent (\cite{Ogura03}; 
Goodwin {\em et al.}~2004; Gruzinov~2005). 
\end{enumerate}

The above show that the problem is well defined inside and
outside the open field light cylinder, and therefore one
can obtain solutions for a general current distribution $A=A(\psi)$.
The two problems are, however, independent, and in general
the solution will be discontinuous at the open field
light cylinder, unless one chooses the one poloidal electric
current distribution $A=A(\psi)$ that will guarantee 
$\psi(r=r_{\rm oflc}^-,z)= \psi(r=r_{\rm oflc}^+,z)$.
Continuity will also result in the smoothness of the
solution (see above boundary condition \#~3).
$A(\psi)$ is obtained as described in CKF
by itteratively correcting to a new function
\[
AA'\left(\psi=\frac{1}{2}\left[\psi(r=r_{\rm oflc}^-,z)
+\psi(r=r_{\rm oflc}^+,z)\right]\right)_{\rm new}
\]
\begin{equation}
=\frac{1}{2}
\left(AA'(\psi(r=r_{\rm oflc}^-,z))_{\rm old}+
AA'(\psi(r=r_{\rm oflc}^+,z))_{\rm old}\right)
\end{equation}
for all grid points along the open field light cylinder.
In the present work
the relaxation inside each grid proceeds
together with the itteration along the open field light cylinder.
This improvement over the CKF method allowed for a
much greater numerical resolution and a much faster
speed of numerical convergence!
Our numerical scheme consists of an elliptic solver
with Chebyshev acceleration (\cite{Press88}) over
two $100\times 100$ numerical grids joined along the open field
light cylinder. 
The discontinuities of $A(\psi)$ and $\Omega_F(\psi)$
accross the separatrix between the open and closed line regions
are smoothed out numerically
over a distance $\delta \psi =0.05$ inside the dead zone.

\section{Steady-state magnetospheric solutions}

\begin{figure}
\includegraphics[angle=270,scale=.50]{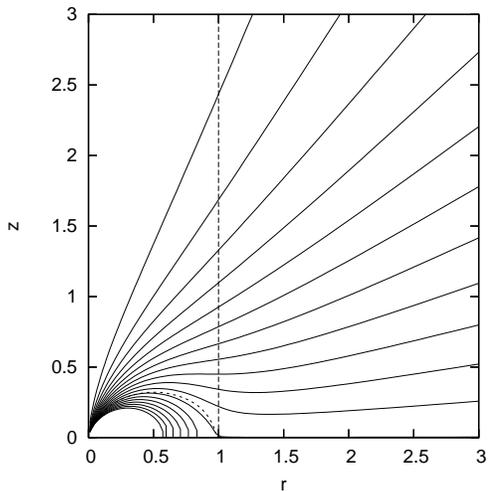}
\caption{$\Omega_F=1$ everywhere (CKF; Gruzinov~2005). 
Thin lines correspond to
$\psi$ intervals of $0.1$. $\psi=0$ along the axis.
The dotted line shows the separatrix
$\psi=\psi_{\rm open}=1.23$. The mathematical singularity is at
$r_{\rm lc}\equiv r_{\rm oflc}=1$.}
\label{figure:1}
\end{figure}

\begin{figure}
\includegraphics[angle=270,scale=.50]{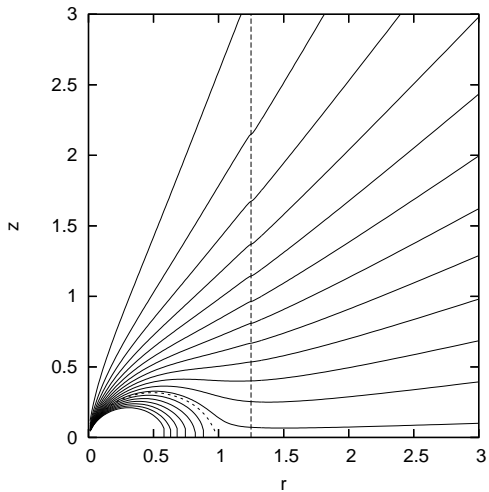}
\caption{$\Omega_{Fo}=0.8$ in the open line region. 
$\psi_{\rm open}=1.23$. $r_{\rm oflc}=1.25$}
\label{figure:2}
\end{figure}

\begin{figure}
\includegraphics[angle=270,scale=.50]{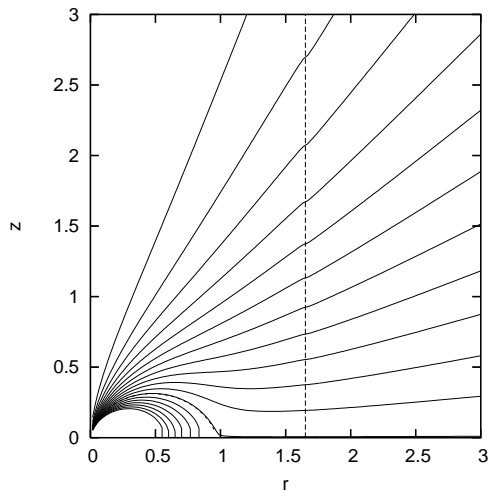}
\caption{$\Omega_{Fo}=0.6$ in the open line region.
Thin lines correspond to
$\psi$ intervals of $0.1$. $\psi=0$ along the axis.
The dotted line shows the separatrix
$\psi=\psi_{\rm open}=1.20$. The mathematical singularity is at
$r_{\rm oflc}=1.67$.}
\label{figure:3}
\end{figure}

\begin{figure}
\includegraphics[angle=270,scale=.50]{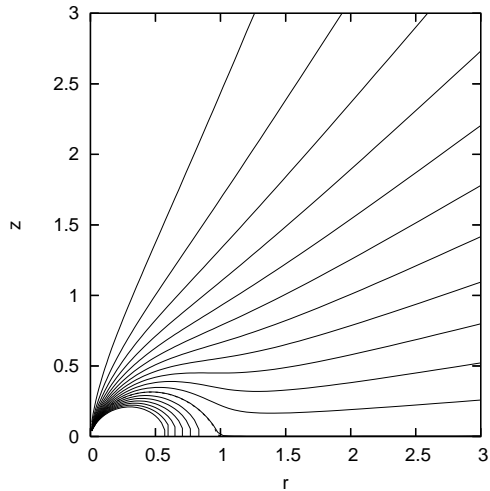}
\caption{$\Omega_{Fo}=0$ in the open line region.
$\psi_{\rm open}=1.20$. $r_{\rm oflc}=\infty$, i.e.
the mathematical singularity is absent in this
limiting case.}
\label{figure:4}
\end{figure}

The various types of solutions of eq.~\ref{pulsareq}
are shown in Figs~\ref{figure:1}-
\ref{figure:4} and Fig.~\ref{figure:8}.
Thin lines correspond to $\psi$ intervals of 0.1. $\psi=0$
along the axis. The
dotted line represents the separatrix $\psi_{\rm open}$.
Fig.~1 shows the CKF case $\Omega_F=1$.
Here, $\psi_{\rm open}=1.23$ within
the accuracy of our numerical simulation.
Figs.~\ref{figure:2} \& \ref{figure:3} show intermediate cases
with $\Omega_{Fo}=0.8$ \& 0.6  
in the open line region respectively. $\psi_{\rm open}=1.23$
\& 1.20 respectively.
Fig.~\ref{figure:4} shows the limiting case with $\Omega_{Fo}= 0$.
In that case there is no light cylinder singularity that
would yield the poloidal electric current distribution $A(\psi)$.
However, we showed previously that 
the poloidal electric current disappears, since it is obtained
as a limit of solutions with
$\Omega_{Fo}\rightarrow 0$ 
in the open line region. Here, $\psi_{\rm open}=1.20$.

The various magnetospheres show a similar poloidal
magnetic field distribution. 
This result is understood since
$B_z$ is approximately $\propto r^{-3}$ in the equatorial
dipole-like closed line region, and therefore an approximate
estimate for $\psi_{\rm open}$ is
\begin{equation}
\psi_{\rm open}\sim \frac{1}{r_c}=1\ .
\end{equation}
However, they
differ significantly in the amount of electric charge and
electric current
they contain in the open field line region, since $\rho_e\propto 
\Omega_{Fo}$ and $A\propto \Omega_{Fo}$. As a result,
they differ in the amount of electromagnetic
field energy they contain in the azimuthal component of the magnetic
field $B_\phi$ and in the electric field $E$, namely
\begin{equation}
\int (B_\phi^2+E^2) r^2 {\rm d}r \sim
\int \left(\frac{\Omega_{F} rB_p}{c}\right)^2
r^2 {\rm d}r \sim
\Omega_{Fo}^2 B_*^2 r_*^3 \left(\frac{r_*}{r_{\rm lc}}\right)^3
\left(\frac{r}{r_{\rm lc}}\right)
\label{energy1}
\end{equation}
Here, the integration distance $r$ extends to distances
$\gg r_{\rm lc}$. Any evolution between the different
solutions will require the release (or buildup) of the corresponding
energy difference (see discussion in the next section).

\begin{figure}
\includegraphics[angle=270,scale=.50]{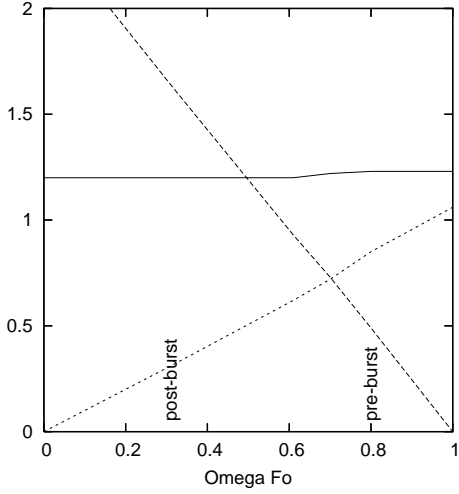}
\caption{Summary of our numerical solutions
applied in the case of SGR~1806-20. We show here
$\psi_{\rm open}$ (continuous line),
the accelerating potential $V_{\rm acc}/(10^{11}\ {\rm Volts})$
(dashed line), and the spindown rate 
$|\dot{P}|/10^{-11}\ {\rm Hz\ sec}^{-1}$ (short dashed line).
On the plot are shown
our estimates for the magnetospheric configuration before
and after the December 27, 2004 burst.}
\label{figure:5}
\end{figure}

We discovered that, as $\Omega_{Fo}$ varies from
$\Omega$ to 0, the open field region decreases
to a minimum value of about $\psi_{\rm open}\sim 1.2$
(see fig.~\ref{figure:5}).
In the next section we will see that this numerical
result might have interesting physical implications
in understanding the SGR phenomenon.

\begin{figure}
\includegraphics[angle=270,scale=.40]{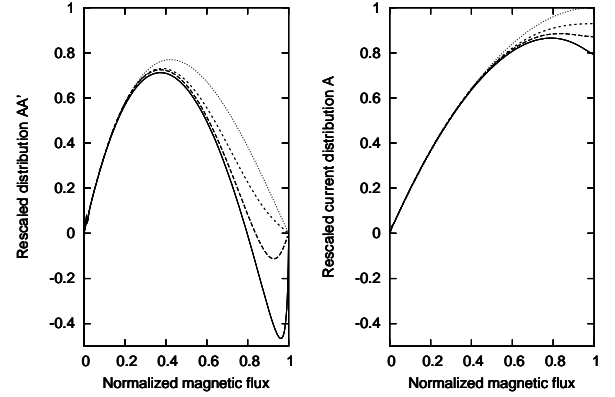}
\caption{The rescaled electric current distribution 
$A/(\psi_{\rm open}\Omega_{Fo})$
and the rescaled distribution
$AA'/(\psi_{\rm open}\Omega_{Fo}^2)$, 
as functions of the rescaled magnetic flux $\psi/\psi_{\rm open}$
in the open line region, for $\Omega_{Fo}=$1, 0.8 \& 0.6 (from
the lower curves up respectively).
The upper curves (dotted) are the ones that correspond to the
Michel split monopole expression.}
\label{figure:6}
\end{figure}

Fig~\ref{figure:6} shows the corresponding rescaled 
electric current distribution $A/(\psi_{\rm open}
\Omega_{Fo})$, and the rescaled distribution
$AA'/(\psi_{\rm open}\Omega_{Fo}^2)$, 
(obtained numerically) as functions of the
normalized magnetic flux $\psi/\psi_{\rm open}$.
\begin{figure}
\includegraphics[angle=270,scale=.40]{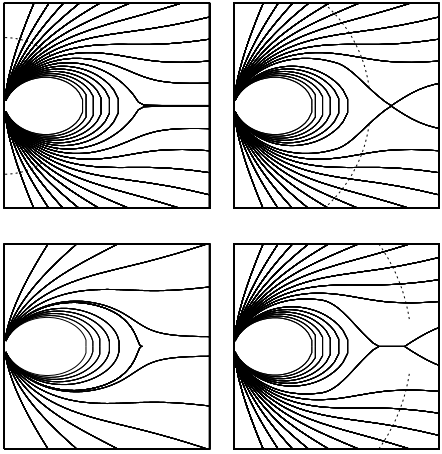}
\caption{Schematic magnetospheric evolution characterized
as `magnetospheric coughing' (clockwise from upper left corner). 
In the upper left corner
is shown a steady-state solution with $\Omega_{Fo}\sim \Omega$.
Conditions at the base of the magnetosphere changed suddenly
towards a different steady-state solution with $\Omega_{Fo}
\neq\Omega$, and a spherical electromagnetic wave
(shown with dotted line)
sweeps through the open field line region at the speed of
light. When the wave reaches
the light cylinder region, reconnection allows the expulsion
of the amount of magnetic flux required for the magnetosphere
to evolve towards the new steady-state solution that corresponds
to the new value of $\Omega_{Fo}$ (upper right corner). 
The detached magnetic flux forms a `plasmoid' that escapes
in the equatorial region (lower right corner). 
The system reaches a final steady-state shown schematically
in the lower left corner, and will remain there for
as long as the physical conditions that sustained the new value
of $\Omega_{Fo}$ at the base of the magnetosphere persist.}
\label{figure:7}
\end{figure}
We see that indeed the electric current
distributions are very similar and proportional to $\Omega_{Fo}$. 
Let us now see how this result affects our estimation of
stellar magnetic fields $B_*$. As we mentioned in the introduction,
it is customary to estimate $B_*$ by equating the observed
stellar spindown energy loss to the estimated electromagnetic
spindown torque. As we show in the Appendix,
\[
L_{\rm em\ spindown}=\Omega\int_{\psi=0}^{\psi_{\rm open}}
A(\psi){\rm d}\psi
\approx \frac{2}{3}\Omega_{Fo}\psi_{\rm open}^2
\]
\begin{equation}
\approx \frac{B_*^2 \Omega^3\Omega_{Fo}r_*^6}{4c^3}
\left(\frac{r_{\rm lc}}{r_c}\right)^2
\label{spindown}
\end{equation}
(in real units).
In general, $r_c$ introduces one more free parameter
in the problem (see section~5). Let us here consider
only the natural case 
$r_c\sim r_{\rm lc}$ and discuss the physical significance
of $\Omega_{Fo}$. Eq.~\ref{spindown}
implies that stellar magnetic 
field estimates need to be revised upwards 
over the canonical value obtained when one
compares eqs.~1 \& 2. Note that when $\Omega_{Fo}= 0$, 
$\rho_e = 0$, $J= 0$, i.e. no currents flow
through the magnetosphere, and therefore the star will not
spin down. In most cases, $\Omega_{Fo}\sim [80,95]\%\Omega$
(Romani, personal communication), and therefore, the correction
introduced in the stellar magnetic field estimate is
in most cases practically insignificant.
The correction is significant and should be taken into 
serious consideration for slow pulsars near the pulsar death-line,
where $V_*(\psi_{\rm open})\approx 10^{12}$~Volts~
$=V(\psi_{\rm open})$ and
$V_F(\psi_{\rm open})\approx 0$ (eq.~\ref{V1}).

\section{A `coughing' magnetosphere}

The solutions presented in the previous section are
all steady-state solutions characterized by one parameter,
$\Omega_{Fo}$, which, as we argued, is determined
by the particle acceleration gap microphysics.
Let's imagine  first that
charge carriers are freely available at the base
of the magnetosphere. 
In that case, the gap is shorted out, and the magnetosphere
is described by a steady-state solution with $\Omega_{Fo}\approx
\Omega$ (CKF). Let us imagine next that 
the supply of charge carriers is somehow suddenly depleted.
The gap will suddenly grow, and the magnetosphere
will quickly evolve towards a different steady-state 
solution with $\Omega_{Fo}\neq \Omega$.
We are now going to discuss how,
in our opinion, the magnetosphere may evolve from
the one steady-state solution to the other. 
We will base our discussion on the particular
example of SGR~1806-20, and its December 27,2004 burst.

We will argue that, when the particle acceleration gap
at the base of the magnetosphere suddenly grows, 
{\em the magnetosphere will spontaneously
evolve from a configuration with a larger open
field line region and a larger poloidal electric current,
to one with a smaller open field line
region and a smaller poloidal electric current}.
One way to achieve this might be through north-south reconnection
at the distance of the light cylinder. 
We expect a significant amount of
magnetic flux ($\sim 5\%\ \psi_{\rm open}$)
to `snap' and move equatorially outward similarly
to a solar coronal mass ejection (plasmoid).
At the same time, the magnetosphere will release the
excess energy contained in the azimuthal component
of the magnetic field $B_\phi$ and in the electric field $E$
through a spherical electromagnetic wave
sweeping through the open field region at the speed of light\footnote{
In general, this will be a spherical Alfven wave moving
outward at the Alfven speed.}.
As is shown in eq.~\ref{energy1}, the energy contained in 
that wave would grow with distance. We would like to characterize this 
dramatic evolution as `magnetospheric coughing'
(see fig.~\ref{figure:7} for a schematic description). 

%
%
%
%

As long as the depletion of charges persists, the
magnetosphere will remain in the low $\Omega_{Fo}$ state.
The magnetosphere might return to 
a higher $\Omega_{Fo}$ state
where angular momentum is removed more efficiently
only if charge carriers become freely available 
again at the base of the magnetosphere.
We speculate that in such case, the magnetosphere
will evolve through differential rotation
between the star and the light cylinder region, and therefore 
the evolution will be less dramatic than the magnetospheric 
coughing described above.

In our example (see fig.~\ref{figure:5}), 
let us choose the solution with $\Omega_{Fo}=0.8$
as the pre-burst solution. Our numerical analysis yielded
\begin{equation}
\Omega_{Fo\ \rm pre-burst}=0.8\ ,
\end{equation}
\begin{equation}
\psi_{\rm open\ pre-burst}=1.23\ .
\end{equation}
Based on our detailed axisymmetric ideal MHD model, and given the
observed pre-burst spindown rate $\dot{P}=-8.5\times 10^{-12}
\ {\rm Hz}/{\rm sec}$, 
we obtain
\begin{equation}
B_*=5.2\times 10^{14}\ {\rm G}\ ,
\end{equation}
and a corresponding 
accelerating potential in the magnetospheric gaps
\begin{equation}
V_{\rm pre-burst}=5\times 10^{10}\ {\rm Volts}\ .
\end{equation}
We know that, after the burst, the spindown rate was 2.7 times
smaller (\cite{Woods05}). This allows us to take
\begin{equation}
\Omega_{Fo\ \rm  post-burst}=0.31\ ,
\end{equation}
\begin{equation}
\psi_{\rm open\ post-burst}=1.20\ .
\end{equation}
\begin{equation}
V_{\rm post-burst}=1.6\times 10^{11}\ {\rm Volts}\ .
\end{equation}
We see that both before and after the burst, the accelerating
potential is of the order of $10^{11}$~Volts.
Indeed, the magnetosphere
is emitting pulsed radiation in both cases, only after
the burst, pulsed radiation is observed to be weaker. We attribute this
difference to the smaller radiation cone (due to the smaller
open field line region) which might thus avoid our line of sight.

According to eq.~\ref{energy1},
the energy difference between the two magnetospheres is of the
order of
\[
(\Omega_{Fo\ \rm pre-burst}^2-
\Omega_{Fo\ \rm post-burst}^2)
B_*^2 r_*^3 \left(\frac{r_*}{r_{\rm lc}}\right)^3
\left(\frac{r}{r_{\rm lc}}\right)
\]
\begin{equation}
\sim 10^{47}\ \left(\frac{r_*}{r_{\rm lc}}\right)^3
\left(\frac{r}{r_{\rm lc}}\right){\rm ergs}\ .
\label{equation2}
\end{equation}
According to \ref{equation2}, the energy contained in the
spherical blast wave will be 
comparable to the apparent burst luminosities observed
on earth (e.g. \cite{Yamazaki05}) 
at distances $r/r_{\rm lc} \sim (r_{\rm lc}/r_*)^3$.
We would like to defer a more detailed discussion of the burst
energetics to a future work.


\section{Conclusions}

In our present work we presented global numerical solutions
of the generalized pulsar equation
that describe the steady-state structure of axisymmetric
rotating neutron star magnetospheres. 
We have introduced 
two new parameters besides the neutron star angular velocity
$\Omega$,
\begin{itemize}
\item $\Omega_{Fo}$, the angular velocity of rotation of
open field lines. This quantity is related to the particle
acceleration gaps at the base of the magnetosphere (the closer
$\Omega_{Fo}$ is to $\Omega$, the smaller the gap), and is determined
by gap microphysics outside the context of our present ideal MHD
formulation.
\item $r_c$, the maximum equatorial extent of the closed line region
(see Appendix).
We speculate that $r_c$ might be determined by inertial effects outside
the context of our present ideal MHD formulation.
\end{itemize}
Note that, in our global solutions,
$\psi_{\rm open}$ (the amount of open field lines)
is determined self-consistently, and consequently it is not
a free parameter (see, however, 
\cite{Goodwin04} for a different point of view).
Similarly to CKF, {\em the} poloidal electric current distribution
that guarantees smoothness and continuity at the open field light cylinder
is obtained itteratively, and an approximate analytic expression is given.
Our results generalize the solution presented in CKF;
\cite{Gruzinov05}.

We also obtained a generalized expression for the 
steady-state spindown magnetospheric energy losses (eq.~\ref{spindown}),
which is different from the canonical one for a misallingned magnetic
rotator. Magnetospheres with different values of $\Omega_{Fo}$
and/or $r_c$ contain different amounts of electric currents, and therefore
spin down differently. This changes slightly our estimates of stellar magnetic
fields $B_*$ (see also \cite{HCK} for a relevant discussion
in the case of magnetar magnetic field estimates). 
More importantly, however, this might have serious
implication in the calculation of the magnetic braking index $n\equiv
\Omega\ddot{\Omega}/\dot{\Omega}^2$. One can easily check 
(eq.~\ref{spindown}) that any functional dependence of
$\Omega_F$ and $\psi_{\rm open}$ {\em different} from the
canonical one $\Omega_F \propto \Omega$, and
$\psi_{\rm open}\propto \Omega$ will yield a braking index
$n\neq 3$ as obtained observationally (Contopoulos \& Spitkovsky, 
in preparation). 

Finally, we argued that the magnetosphere may spontaneously
evolve between steady-state  configurations characterized by different
values of $\Omega_{Fo}$ and/or $r_c$. The evolution from
a high to low value of $\Omega_{Fo}$ 
and/or low to high value of $r_c$ will result in the dramatic
release of a significant amount of electromagnetic field
energy and magnetic flux. The return to the former configuration
will be less dramatic, since it will require the
buildup of the corresponding electromagnetic field energy
difference. Our results might be relevant in understanding 
the SGR burst phenomenon.

\acknowledgements{We would like to thank
Christos Eftymiopoulos and Demos Kazanas 
for their support in reviving this intriguing problem.
We would also like to thank Jonathan Arons,
Roger Blandford, Roger Romani, and Anatoly Spitkovsky
for interesting discussions and comments. We would finally like to
acknowledge the contribution of the unknown referee
in improving the presentation of our ideas.}

\appendix

\section{Pulsar spindown estimates}

When a neutron star is not surrounded by vacuum,
the rotating charged relativistic Goldreich-Julian-type 
magnetosphere is threaded by poloidal and toroidal electric currents.
We will consider only the axisymmetric case for simplicity.
Two large scale poloidal electric current circuits
(north \& south) are generated. These flow only along
open field lines, and
close along the surface of the star at the two polar caps where
they generate electromagnetic torques antiparallel
to the angular momentum of the neutron star
\begin{equation}
\frac{1}{c}r B J {\rm d}S{\rm d}r
\label{sc}
\end{equation}
through any stellar cross section ${\rm d}S$ threaded
by poloidal electric current density $J$. One can easily check
that the stellar kinetic energy loss
through the above torques is given by
\begin{equation}
L_{\rm em\ spindown}=\Omega\int_{\psi=0}^{\psi_{\rm open}}
A(\psi){\rm d}\psi\simeq
\frac{2}{3}\Omega\Omega_{Fo}\psi_{\rm open}^2\approx \Omega_{Fo}
\label{sdint}
\end{equation}
(our expression accounts for the two hemispheres, north \& south).
We made use of the numerical result $\psi_{\rm open}\approx 1.23$.
At the same time, the magnetosphere radiates electromagnetic
energy 
\begin{equation}
\frac{c}{4\pi}r E_p B_\phi {\rm d}S
\label{em}
\end{equation}
through any cross section ${\rm d}S$ in the
region of open field lines. One can easily check
that the total electromagnetic energy loss
through the above Poynting flux is given by
\begin{equation}
L_{\rm em}=\int_{\psi=0}^{\psi_{\rm open}}
A(\psi)\Omega_F(\psi){\rm d}\psi\simeq
\frac{2}{3}\Omega_{Fo}^2\psi_{\rm open}^2
\approx \Omega_{Fo}^2\ .
\label{emint}
\end{equation}
$\Omega_F$ is in general smaller than $\Omega$, and therefore,
$L_{\rm em}$ is in general less than $L_{\rm em\ spindown}$.
The difference between the two is consumed in the 
particle acceleration gaps that develop along open
field lines, namely
\[
L_{\rm particles}=L_{\rm em\ spindown}-L_{\rm em}
=\int_{\psi=0}^{\psi_{\rm open}}
A(\psi)(\Omega-\Omega_F(\psi)){\rm d}\psi
\]
\begin{equation}
\simeq \frac{2}{3} \Omega_{Fo}(1-\Omega_{Fo})\psi_{\rm open}^2
\approx \Omega_{Fo}(1-\Omega_{Fo})\ .
\end{equation}
The above expressions are normalized to the
Goldreich-Julian value
\begin{equation}
L_{\rm GJ}\equiv \frac{B_*^2 \Omega^4 r_*^6}{4c^3}\ .
\end{equation}

\section{Alternative magnetospheric solutions}

\begin{figure}
\includegraphics[angle=270,scale=.50]{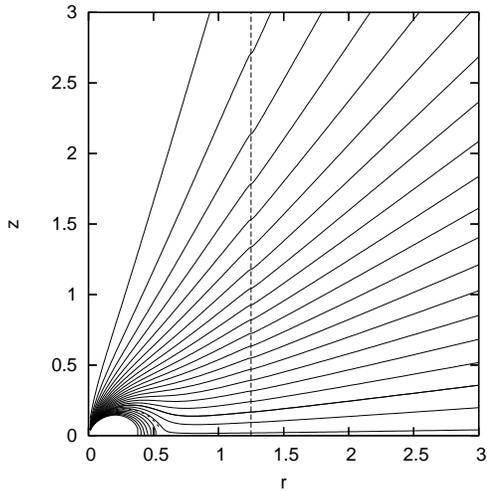}
\caption{A solution with 2.7 times more efficient spindown
than the solution shown in fig.~\ref{figure:2}. 
The `dead zone' extends up to $r_{\rm c}=0.59$.
$\Omega_{Fo}=0.8$. $\psi_{\rm open}=2.03$.}
\label{figure:8}
\end{figure}

In solving eq.~\ref{pulsareq},
we have all along argued that nature will choose the most natural
solution, namely the one with the maximum extent of the `dead zone'.
A competing to the above scenario might be one where
the extent of the `dead zone' is
a free parameter $r_{\rm c}$ (Goodwin {\em et al.}~2004).
Since $\Omega_{F}$ introduces one more free parameter
in the problem, we will consider only one representative
case with $\Omega_{Fo}=0.8$. Eq.~\ref{pulsareq}
can be solved numerically as described before.
In this scenario, solutions with
a smaller `dead zone' are also more efficient in removing
angular momentum from the spinning star (eq.~\ref{sdint}). 
As an example, we take $r_{\rm c}\sim 0.59$ and obtain
$\psi_{\rm open}=2.03$ (fig.~\ref{figure:8}). 
This solution may evolve rapidly
through reconnection towards the solution shown 
in fig.~\ref{figure:2}
with $r_{\rm c}\sim 1$, $\psi_{\rm open}=1.23$,
and thus yield a spindown rate 2.7 times lower, releasing
at the same time a significant amount of magnetic field
energy. Note that the system is an efficient
radiator through particle acceleration processes both
before and after the burst ($\Omega_{Fo}<1$). 


\begin{thebibliography}{}
\bibitem[Beskin 1997]{Beskin97}
Beskin, V.S. 1997, Physics-Uspekhi, 40, 659
\bibitem[Beskin \& Malyshkin 1998]{BeskinM}
Beskin, V.S. \& Malyshkin, L. M. 1998, MNRAS, 298, 847
\bibitem[Contopoulos, Kazanas, Fendt 1999]{CKF}
Contopoulos, I., Kazanas, D. \& Fendt, C. 1999, ApJ, 511, 351
\bibitem[Goldreich \& Julina 1969]{GJ69}
Goldreich, P. \& Julian, W. H. 1969, ApJ, 157, 869
\bibitem[Goodwin {\em et al.} 2004]{Goodwin04}
Goodwin, S. P., Mestel, J., Mestel, L. \& Wright, G. A. E. 2004,
MNRAS, 349, 213
\bibitem[Gruzinov 2005]{Gruzinov05}
Gruzinov A. 2005, Phys. Rev. Lett., 94
\bibitem[Harding, Contopoulos \& Kazanas 1999]{HCK}
Harding, A. K., Contopoulos, I. \& Kazanas, D. 1999, ApJ, 525, 125
\bibitem[Hibschman \& Arons 2001]{HArons01}
Hibschman, J. A. \& Arons, J. 2001, ApJ, 554, 624
\bibitem[Lyubarskii 1990]{Lyubarskii90}
Lyubarskii, Y. E. 1990, Sov. Astron. Lett., 16,16
\bibitem[Michel 1974]{Michel74}
Michel, F. C. 1974, ApJ, 192, 713
\bibitem[Ogura \& Kojima 2003]{Ogura03}
Ogura, J. \& Kojima, Y. 2003, Progr. of Theor. Phys., 109, 619
\bibitem[Palmer {\em et al.} 2005]{Palmer05}
Palmer, D. M. {\em et al.}, 2005, astro-ph/0503030
\bibitem[Press {\em et al.} 1988]{Press88}
Press, W. H., Flamery, B. P., Teukolsky, S. A. \& Vetterling, W. T.
1988, Numerical Recipes (Cambridge Univ. Press: Cambridge)
\bibitem[Rankin \& Wright 2003]{Rankin03}
Rankin, J. M. \& Wright, G. A. E. 2003, A\&A Rev., 12, 43
\bibitem[Ruderman \& Sutherland 1975]{RS75}
Ruderman, M.A. \& Sutherland, P. G. 1975, ApJ, 196, 51
\bibitem[Spitkovsky 2004]{Spitkovsky04}
Spitkovsky, A. 2004, IAUS, 218, 357
\bibitem[Uzdensky 2003]{Uzdensky03}
Uzdensky, D. A. 2003, ApJ, 598, 446
\bibitem[Woods {\rm et al.} 2005]{Woods05}
Woods, P. M. {\em et al.} 2005, Astronomer's Telegram No 407
\bibitem[Yamazaki {\em et al.} 2005]{Yamazaki05}
Yamazaki R., Ioka, K., Takahara, F. \& Shibazaki, N. 2005, 
astro-ph/0502320
\end{thebibliography}
\end{document}